\documentclass[pdftex]{aastex61}
\usepackage{graphicx}

\shorttitle{Substructures and distortions in the Magellanic periphery}
\shortauthors{Mackey et al.}

\begin{document}

\title{Substructures and tidal distortions in the Magellanic stellar periphery}

\author{Dougal Mackey}
\affiliation{Research School of Astronomy and Astrophysics, Australian National University, Canberra, ACT 2611, Australia}

\author{Sergey E. Koposov}
\affiliation{McWilliams Center for Cosmology, Department of Physics, Carnegie Mellon University, 5000 Forbes Avenue, Pittsburgh, PA 15213, USA}
\affiliation{Institute of Astronomy, University of Cambridge, Madingley Road, Cambridge, CB3 0HA, UK}

\author{Gary Da Costa}
\affiliation{Research School of Astronomy and Astrophysics, Australian National University, Canberra, ACT 2611, Australia}

\author{Vasily Belokurov}
\affiliation{Institute of Astronomy, University of Cambridge, Madingley Road, Cambridge, CB3 0HA, UK}
\affiliation{Center for Computational Astrophysics, Flatiron Institute, 162 5th Avenue, New York, NY 10010, USA}

\author{Denis Erkal}
\affiliation{Department of Physics, University of Surrey, Guildford, GU2 7XH, UK}

\author{Pete Kuzma}
\affiliation{Research School of Astronomy and Astrophysics, Australian National University, Canberra, ACT 2611, Australia}
\affiliation{Institute for Astronomy, University of Edinburgh, Royal Observatory, Blackford Hill, Edinburgh, EH9 3HJ, UK}

\begin{abstract}
We use a new panoramic imaging survey, conducted with the Dark Energy Camera, to map the stellar fringes of the Large and Small Magellanic Clouds to extremely low surface brightness $V\ga32$ mag arcsec$^{-2}$. Our results starkly illustrate the closely interacting nature of the LMC-SMC pair. We show that the outer LMC disk is strongly distorted, exhibiting an irregular shape, evidence for warping, and significant truncation on the side facing the SMC. Large diffuse stellar substructures are present both to the north and south of the LMC, and in the inter-Cloud region. At least one of these features appears co-spatial with the bridge of RR Lyrae stars that connects the Clouds. The SMC is highly disturbed -- we confirm the presence of tidal tails, as well as a large line-of-sight depth on the side closest to the LMC.  Young, intermediate-age, and ancient stellar populations in the SMC exhibit strikingly different spatial distributions. In particular, those with ages $\sim1.5-4$ Gyr exhibit a spheroidal distribution with a centroid offset from that of the oldest stars by several degrees towards the LMC. We speculate that the gravitational influence of the LMC may already have been perturbing the gaseous component of the SMC several Gyr ago. With careful modeling, the variety of substructures and tidal distortions evident in the Magellanic periphery should tightly constrain the interaction history of the Clouds.
\end{abstract}


\section{Introduction}
As the two largest satellites of the Milky Way, and amongst the closest, the Large and Small Magellanic Clouds (LMC/SMC) are important to a host of contemporary problems in near-field astrophysics. Not only is their presence statistically unusual \citep[e.g.,][]{tollerud:11,robotham:12}, it has significant implications for measurements of the Milky Way's dark halo \citep{veraciro:13,gomez:15} and for interpreting the number and distribution of faint Galactic satellites \citep{jethwa:16,sales:17}. Moreover, the Clouds are responsible for depositing $\ga5\times10^8\,{\rm M_\odot}$ of neutral hydrogen gas into the Milky Way's outskirts, forming the Magellanic Bridge and the $\approx200\degr$ long Magellanic Stream \citep{bruns:05,nidever:10}.\vspace{2mm}

The orbits and interaction histories of the LMC and SMC are central to our understanding of these issues. Precise proper motion measurements have revealed that the Clouds must either be on their first pericentric passage about the Milky Way or on a very long period ($\ga4$ Gyr) orbit, where the uncertaintly is largely due to the poorly-known masses of the LMC and our Galaxy \citep{kallivayalil:13}. In either scenario the stripping of gas to form the Magellanic Bridge and Stream is best explained by repeated close interactions between the Clouds \citep{besla:10,besla:12,diaz:12}. The number, timing, and severity of these events is not well constrained, but they are likely also responsible for many other peculiar characteristics of the system, including the Clouds' bursty, coupled star-formation histories \citep{harris:09}, the irregular morphology and internal kinematics of the SMC \citep{nidever:13,belokurov:17,zivick:18}, the off-centre stellar bar in the LMC \citep{vdm:01b}, and the presence of SMC stars in the inter-Cloud region \citep{noel:15,carrera:17,belokurov:17} and LMC outskirts \citep{olsen:11,deason:17}.\vspace{2mm}

Mapping stellar structures in the periphery of the Clouds is a potentially powerful, but only recently feasible, means of unraveling the Magellanic interaction history. Stars in the outskirts of the system are weakly bound and hence strongly susceptible to external perturbations, the signatures of which may persist for long periods. Pioneering studies showed that the LMC and SMC extend to surprisingly large radii $\sim15\degr-20\degr$, but yielded conflicting results regarding their overall shape and dimensions \citep[e.g.,][]{munoz:06,majewski:09,saha:10}. More recently, \citet{mackey:16} used deep contiguous imaging from the Dark Energy Survey (DES) to uncover a striking arc-like substructure in the northern LMC outskirts, which \citet{besla:16} showed could plausibly arise from repeated close encouters between the Clouds, while \citet{belokurov:17} used Gaia DR1 photometry to reveal a bridge of stellar debris connecting the LMC and SMC -- apparently yet another imprint of the Magellanic interaction history.\vspace{2mm}

In this Letter we present initial results from a new panoramic survey of the Magellanic stellar periphery, designed to obtain the first global view of the system at extremely low surface brightness and explore the overall extent and morphology of structural distortions in the outer regions of the Clouds.\vspace{2mm}

\section{Observations and data analysis}
We utilise a photometric catalog derived from two sources: the first DES public data release (DES-DR1), and our own survey observations. All imaging was obtained using the Dark Energy Camera \citep[DECam;][]{flaugher:15}, which is a 520 megapixel imager with a $\sim3$\ deg$^2$ field-of-view mounted on the 4m Blanco Telescope at the Cerro Tololo Inter-American Observatory in Chile. The DES-DR1 source catalog was produced as described in papers accompanying the data release \citep{abbott:18,morganson:18}. Our observations were conducted over two separate runs: four first-half nights on 2016 February 25-28, and two full nights on 2017 October 7-8 (programs 2016A-0618 and 2017B-0906, PI: Mackey). This imaging uses the $g$ and $r$ filters, and covers $\approx440$\ deg$^2$ spanning the western and southern outskirts of the LMC, the entire inter-Cloud region, most of the SMC periphery, and an area adjoining the DES footprint north-east of the LMC.\vspace{2mm}

Our raw data were processed by the DECam community pipeline \citep{valdes:14}. We carried out source detection and photometric measurements using the {\tt SExtractor} and {\tt PSFEx} software \citep{bertin:96,bertin:11}, and then constructed a point-source catalogue by merging individual detection lists and excluding galaxies using the {\sc spread\_model} and {\sc spreaderr\_model} parameters as in \citet{koposov:15}. To calibrate our photometry we employed DR9 of the APASS survey after removing the presence of small-scale systematics \citep[see][]{koposov:18} and transforming into the DES-DR1 AB system. Small regions of overlap between our imaging and DES-DR1 exhibit offsets $\la0.03$ mag. We corrected all photometry for foreground reddening using the prescription of \citet{schlafly:11}.

\section{Results}
\begin{figure*}
\begin{center}
\includegraphics[height=88.5mm]{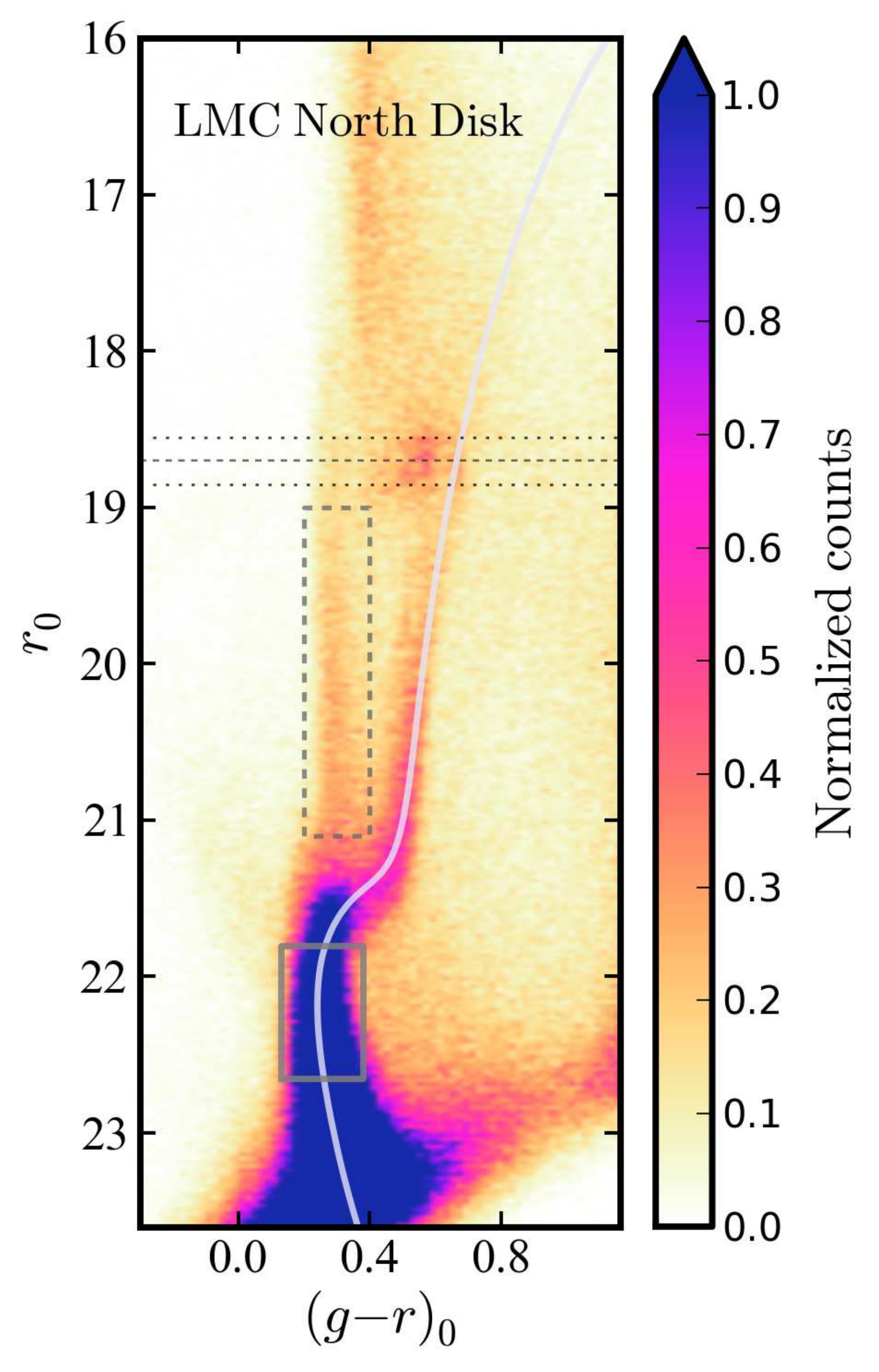}
\hspace{0mm}
\includegraphics[height=88mm]{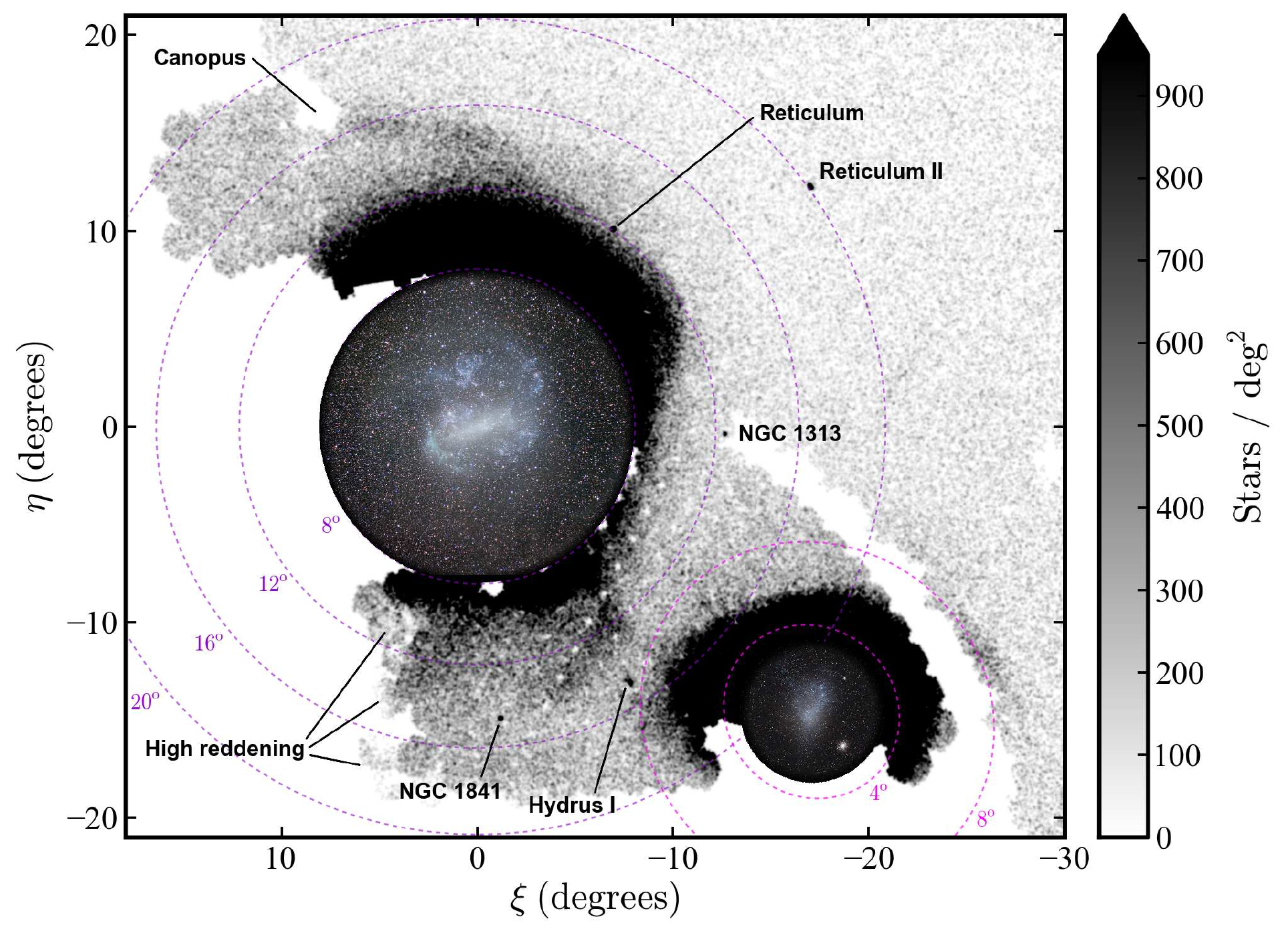}
\end{center}
\caption{{\bf Left:} Hess diagram for stars in the outer northern LMC disk, between radii $10\degr-12\degr$ and position angles $\pm30\degr$ of north. The white line denotes a MIST isochrone with age $11$ Gyr and $[$Fe$/$H$]=-1.3$ shifted to a distance modulus $\mu=18.45$. Our MSTO selection box is marked (solid), as is a selection box for the Milky Way halo (dashed). Dotted lines illustrate the red clump level and spread. {\bf Right:} Map showing the surface density of Magellanic MSTO stars. This is a tangent plane projection centred on the LMC; at the origin, north is up and east is to the left. The dashed circles mark various angular separations from the LMC and SMC, as indicated.
\label{f:cmdmap}}
\end{figure*}

In the left-hand panel of Figure \ref{f:cmdmap} we show a Hess diagram for all stars lying between $10\degr-12\degr$ from the LMC and within position angles $\pm30\degr$ of north. The stellar population of the outer LMC disk is prominent and is well described by a MIST isochrone \citep{choi:16,dotter:16} of age $11$ Gyr and $[$Fe$/$H$]=-1.3$, shifted to a distance modulus $\mu=18.45$. A distinct red clump is also visible.\vspace{2mm}

We use this color-magnitude diagram (CMD) as a template to define a selection box around the main-sequence turn-off (MSTO) that will allow us to map the spatial density of ancient stars across the Magellanic system with high contrast relative to contaminants. We elected not to employ a formal matched-filtering scheme due to the significant variations in stellar population characteristics and line-of-sight distance in the Clouds; the color bounds of our selection box, and its bright limit, were chosen empirically to minimize sensitivity to such variations. The faint limit is set to ensure robustness against spatial variations in detection completeness (see below).\vspace{2mm}

Our spatial density map is displayed in the right-hand panel of Figure \ref{f:cmdmap}, and reveals striking overdensities and structural distortions in the Magellanic periphery. North of the LMC is the arc-like feature discovered by \citet{mackey:16}, which extends behind the bright star Canopus to a radial distance $\approx20\degr$. To the west and south the outer LMC disk appears substantially truncated, and in the far south two large, previously-unknown substructures are visible. The SMC is significantly elongated in the direction of these substructures, and on the anti-LMC side the overdensity discovered by \citet{pieres:17} is evident at radius $\approx8\degr$. An area of high reddening affects the edge of the survey region to the south-east of the LMC. Based on calculations by \citet{mackey:16}, our map is sensitive to features with $V$-band surface brightness as faint as $\ga32$\ mag\ arcsec$^{-2}$.\vspace{2mm}
 
Several compact overdensities are also apparent. These include the LMC globular clusters NGC~1841 and Reticulum, the background galaxy NGC~1313, the ultra-faint dwarf galaxy Reticulum~II \citep{koposov:15,bechtol:15}, and a new ultra-faint dwarf projected in the inter-Cloud region, Hydrus~I, which is the focus of a companion paper \citep{koposov:18}.\vspace{2mm}

\begin{figure*}
\begin{center}
\includegraphics[height=64mm]{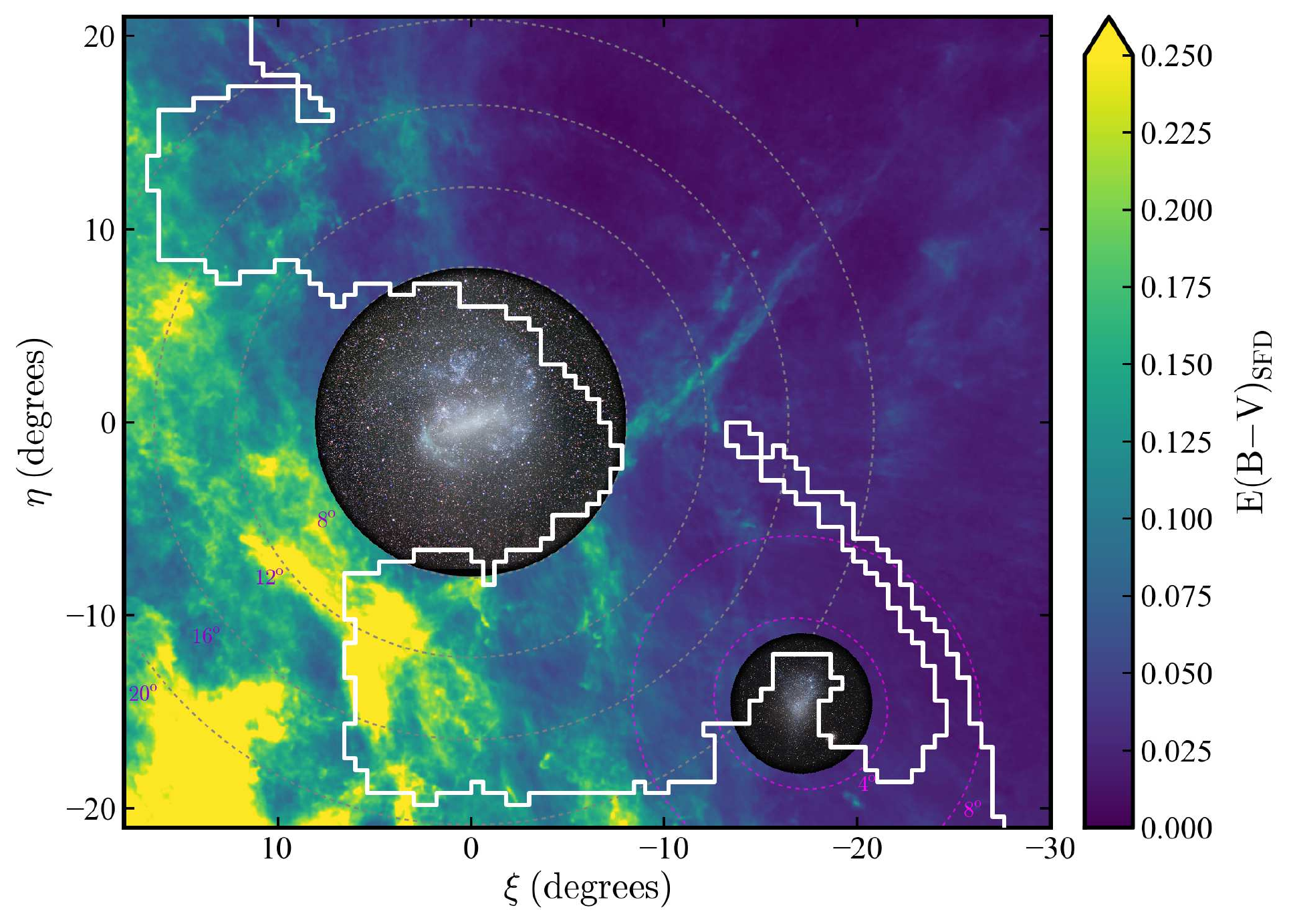}
\hspace{0mm}
\includegraphics[height=64mm]{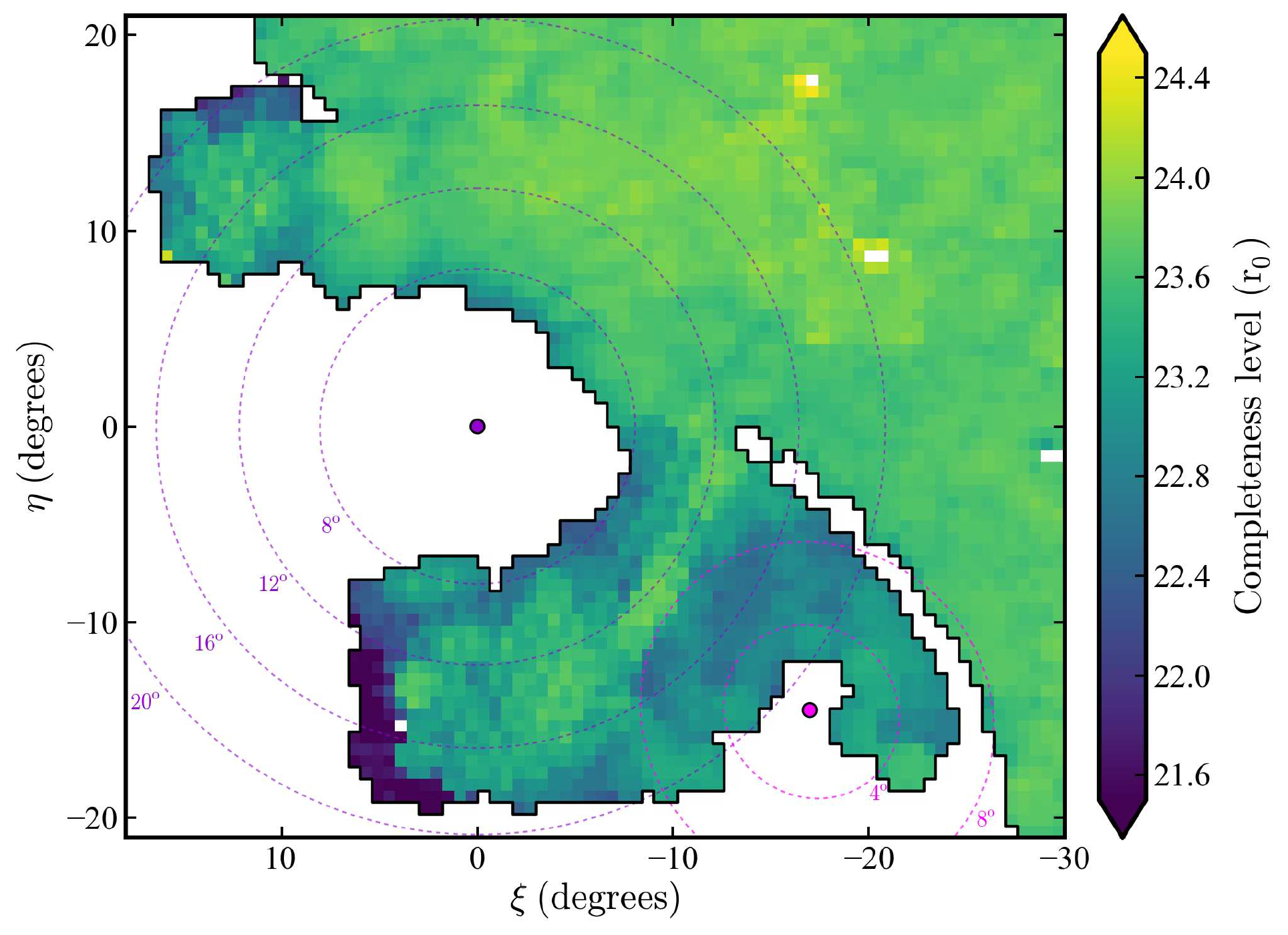}\\
\vspace{1mm}
\includegraphics[height=64mm]{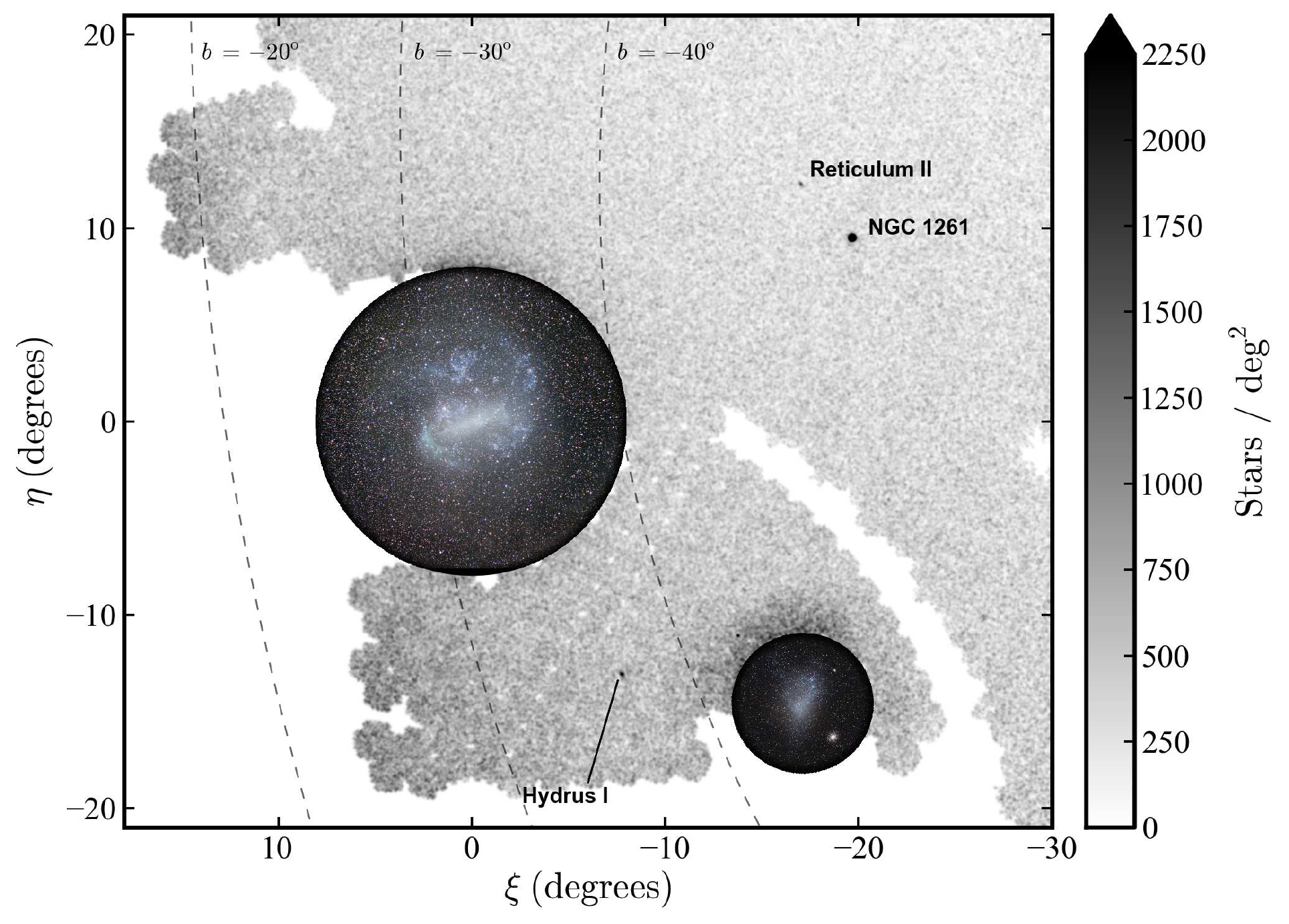}
\hspace{0mm}
\includegraphics[height=64mm]{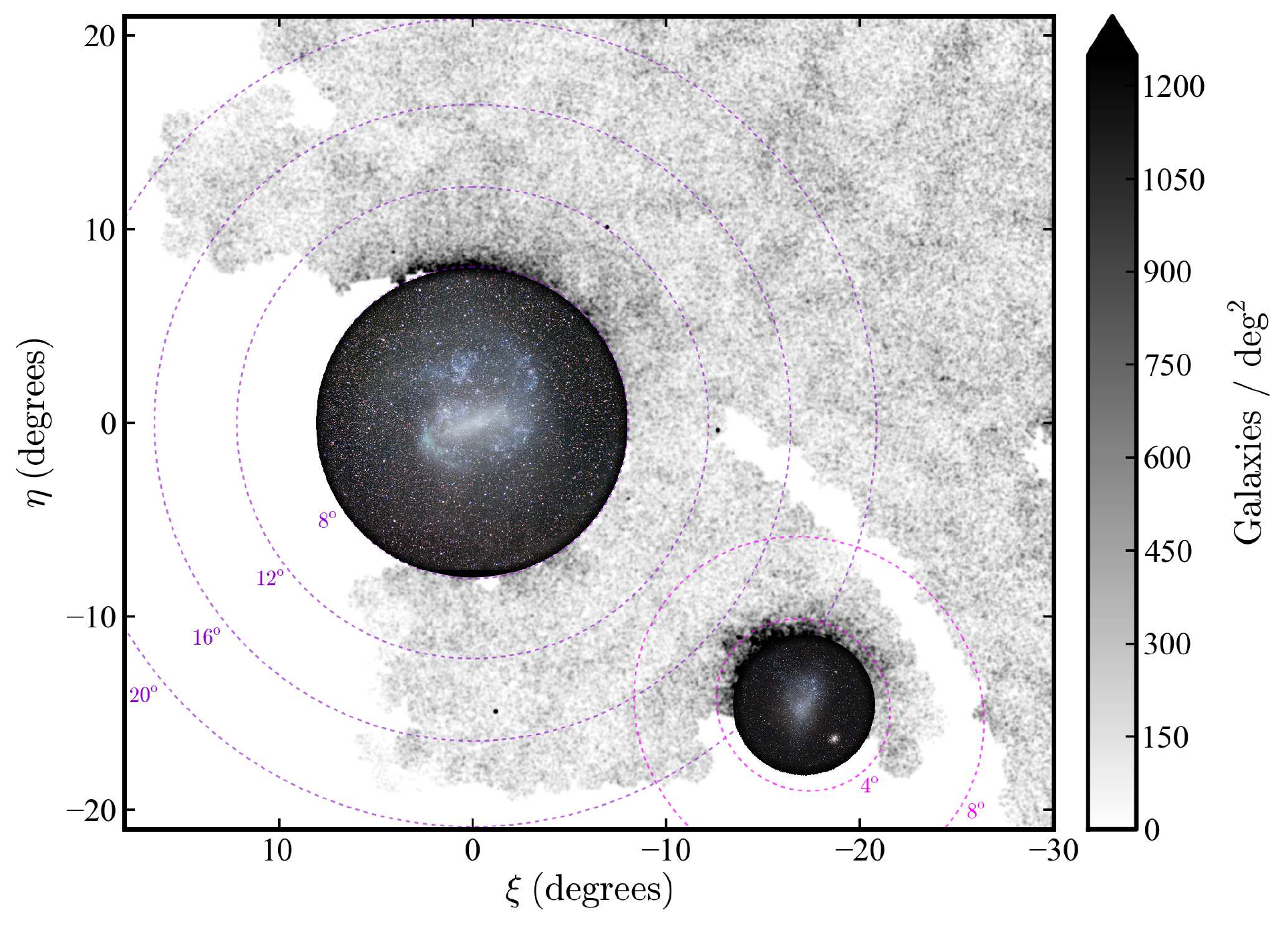}
\end{center}
\caption{Ancillary maps showing the spatial variation of interstellar reddening (upper-left), detection completeness (upper-right), foreground halo populations (lower-left), and background galaxies (lower-right) across the survey footprint. Reddening values are from \citet{schlegel:98}, while the completeness limits are calculated in $0.6\degr\times0.6\degr$ bins (see text). Halo populations are selected using the box defined in Figure \ref{f:cmdmap}; on the density map for these stars, different values of the Galactic longitude are marked.
\label{f:ancillary}}
\end{figure*}
 
In Figure \ref{f:ancillary} we display four ancillary maps to demonstrate that the structures evident in Figure \ref{f:cmdmap} are not due to observational artefact. The top-left panel shows that foreground reddening is generally low across the survey area, with $E(B-V)_{\rm SFD}\la0.15$ except in the aforementioned region south-east of the LMC. The top-right panel charts the de-reddened $r$-band magnitude where our stellar catalogue begins to become significantly affected by incompleteness. We define this in $0.6\degr\times0.6\degr$ bins by locating the turn-over in the luminosity function for stars with $-0.2\leq(g-r)_0\leq0.8$ and fitting a curve to determine the level at which the number counts fall to $75\%$ of this peak. We imposed the color cut to capture the behavior of ancient MSTO-like stars. The DES region (covering the upper-right half of the map) has a rather uniform completeness level near $r_0\approx23.8$; in contrast the area covered by our survey is shallower, and, due to variable weather conditions and lunar illumination, somewhat patchier. However, excluding the small region strongly affected by reddening, our incompleteness limit is always fainter than $r_0\sim22.8$.\vspace{2mm}

The foreground populations closest to Magellanic MSTO stars on the CMD are due to the Milky Way halo. The lower-left panel of Figure \ref{f:ancillary} maps the density of these contaminants selected using the dashed box defined in Figure \ref{f:cmdmap}. The distribution is quite homogeneous with only a mild increase towards lower Galactic latitudes, likely due to distant thick-disk stars. Minor contamination near the SMC from intermediate-age subgiants is also visible (see below). The lower-right panel of Figure \ref{f:ancillary} shows the spatial density of sources in our MSTO box classified as galaxies. While these are not the {\it unresolved} contaminants that could affect our primary map, they ought to be distributed in a similar fashion. This distribution exhibits only a weakly fluctuating filamentary structure.\vspace{2mm}

\begin{figure*}
\begin{center}
\includegraphics[height=68mm]{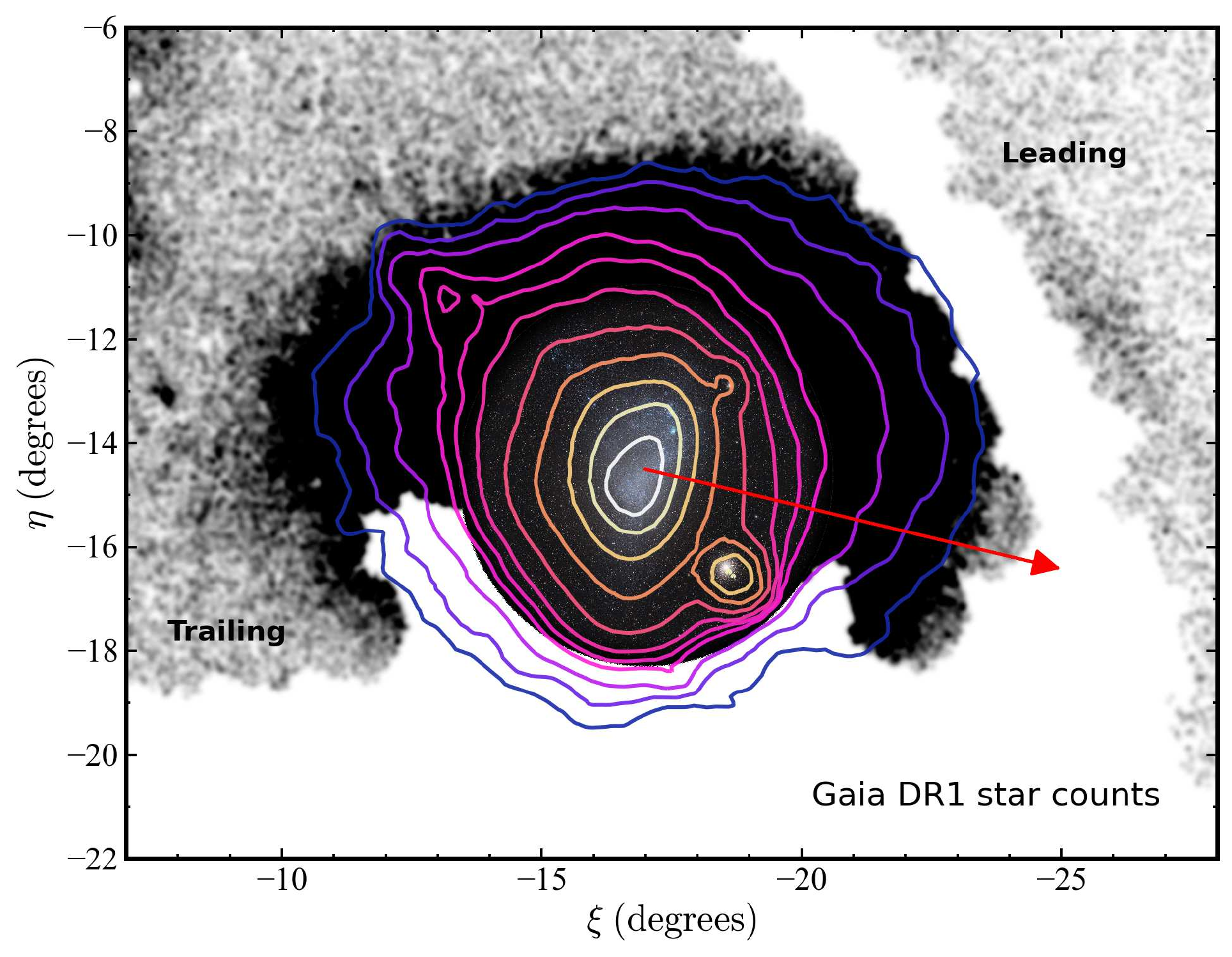}
\hspace{0mm}
\includegraphics[height=68mm]{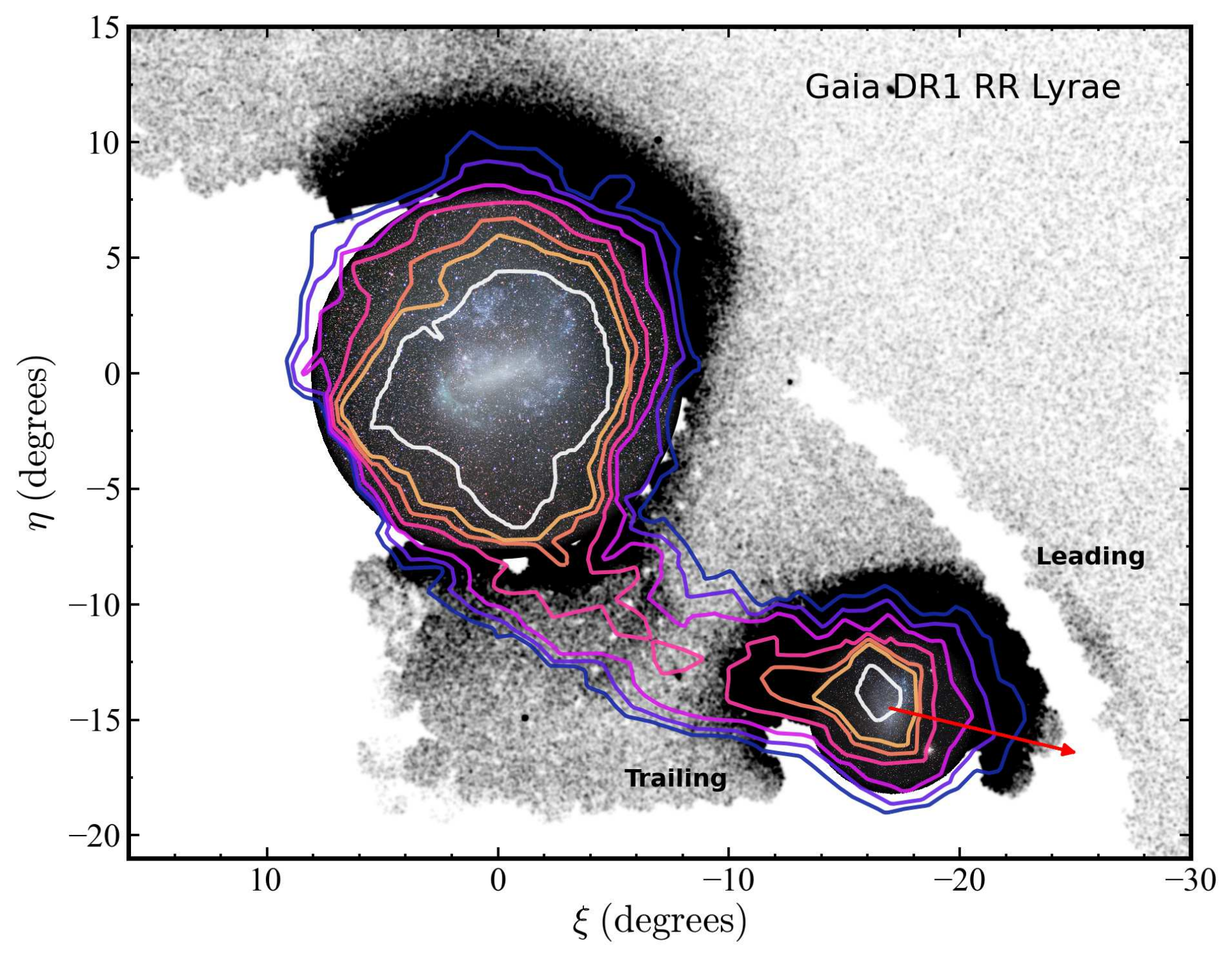}
\end{center}
\caption{Gaia DR1 density contours from \citet{belokurov:17} over-plotted on our DECam map. {\bf Left:} SMC star counts, showing indications of tidal distortion. The contours span $20-2\times10^5$ deg$^{-2}$ at logarithmic intervals. The direction of the SMC's proper motion relative to the LMC is indicated (red arrow). Note the Galactic globular clusters 47~Tuc and NGC~362 superimposed on the SMC interior, and the SMC ``wing'' extending to the upper left. {\bf Right:} RR Lyrae star counts, showing the narrow bridge connecting the two Clouds. The contours span $5-100$ deg$^{-2}$ at logarithmic intervals. 
\label{f:gaia}}
\end{figure*}
 
Having established the robustness of the features visible in our map of the Magellanic system, we turn to a disucssion of their properties. Using Gaia DR1 photometry, \citet{belokurov:17} found that (i) the outer density contours of the SMC appear to indicate significant tidal distortion, and (ii) there is a bridge of RR Lyrae stars connecting the two Clouds. In Figure \ref{f:gaia} we overplot these authors' density contours on our map. While the Gaia SMC counts do not probe as faint as does our survey, their outer levels closely match the DECam density distribution and they provide a clearer view of the SMC's central regions. The twisting of the Gaia isophotes into S-shaped tidal tails noted by \citet{belokurov:17} is evident. Our map shows that the overdensity discovered by \citet{pieres:17} is clearly an extension of the Belokurov et al.\ leading tail. On the opposite side the RR Lyrae bridge joins the SMC at the base of the trailing tail. The overall elongation of the SMC is aligned with both this bridge and the proper motion of the SMC relative to the LMC. The prominent inter-Cloud overdensity near $(\xi,\,\eta)\sim(-6,\,-11)$ in our map appears co-spatial with the RR Lyrae bridge, and may be associated with this larger structure. Notably, however, the southern overdensity near $(2,\,-12)$ falls outside the RR Lyrae contours.\vspace{2mm}

\begin{figure*}
\begin{center}
\includegraphics[height=78mm]{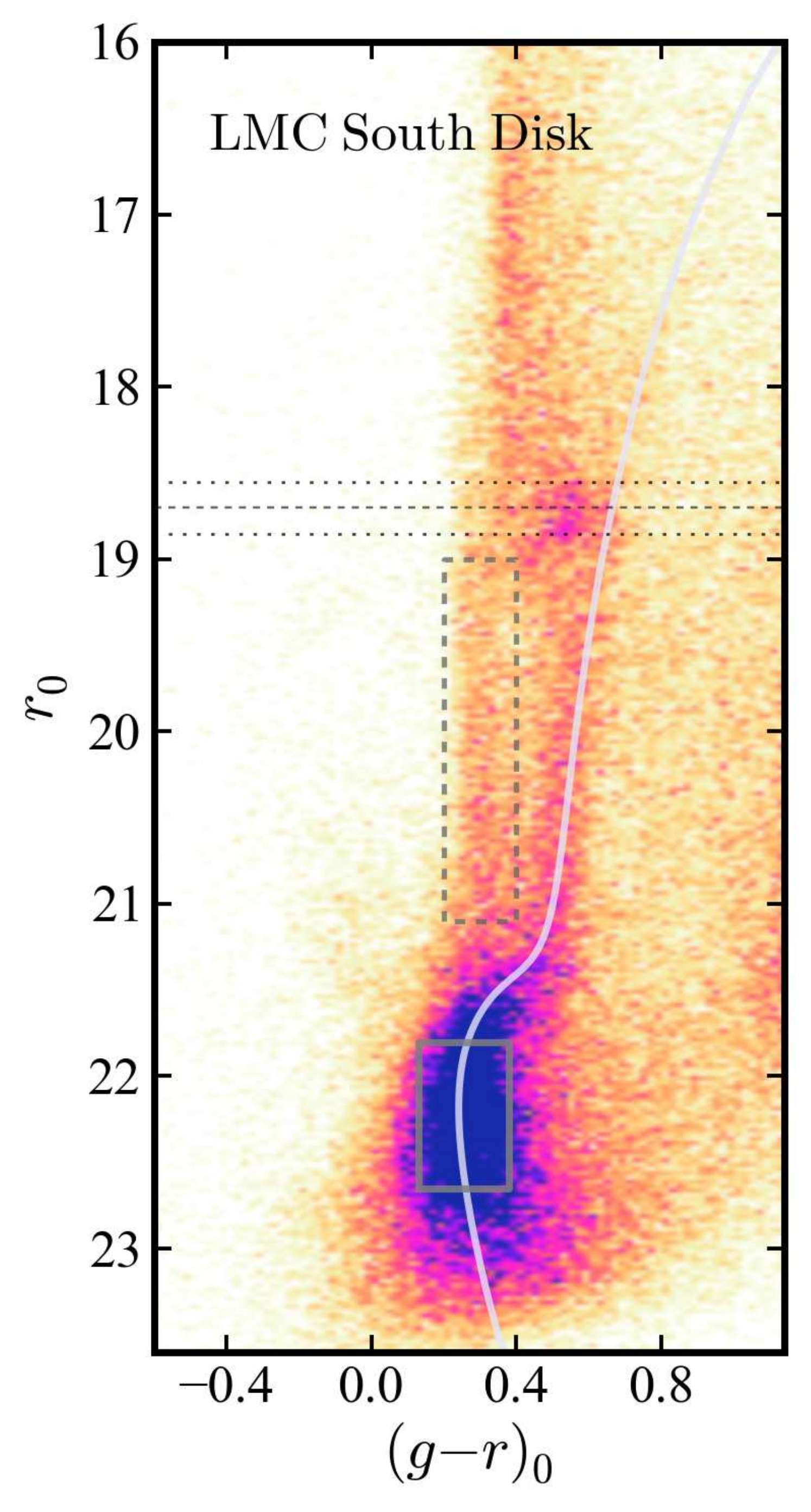}
\hspace{0mm}
\includegraphics[height=75mm]{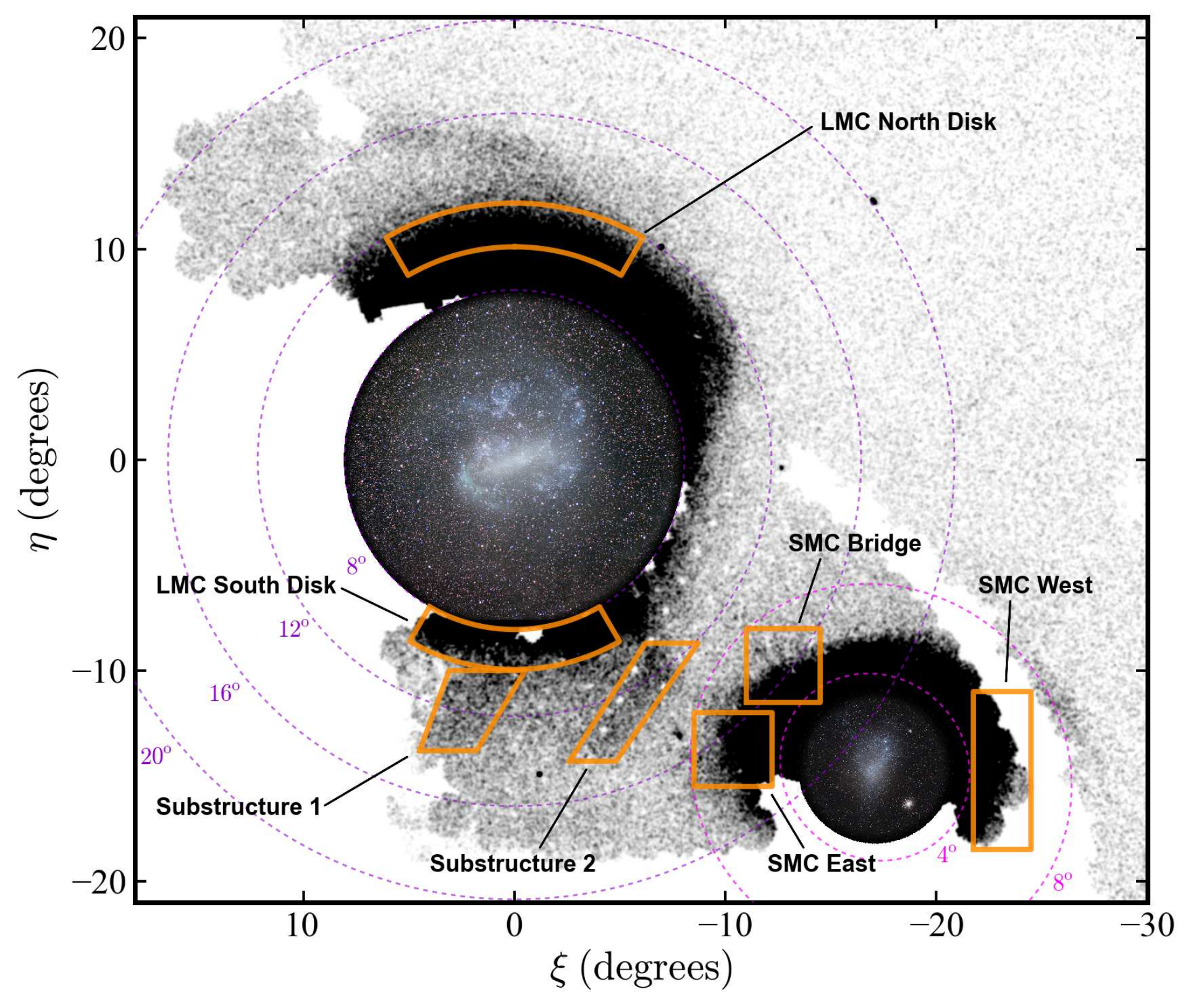}
\hspace{0mm}
\includegraphics[height=78mm]{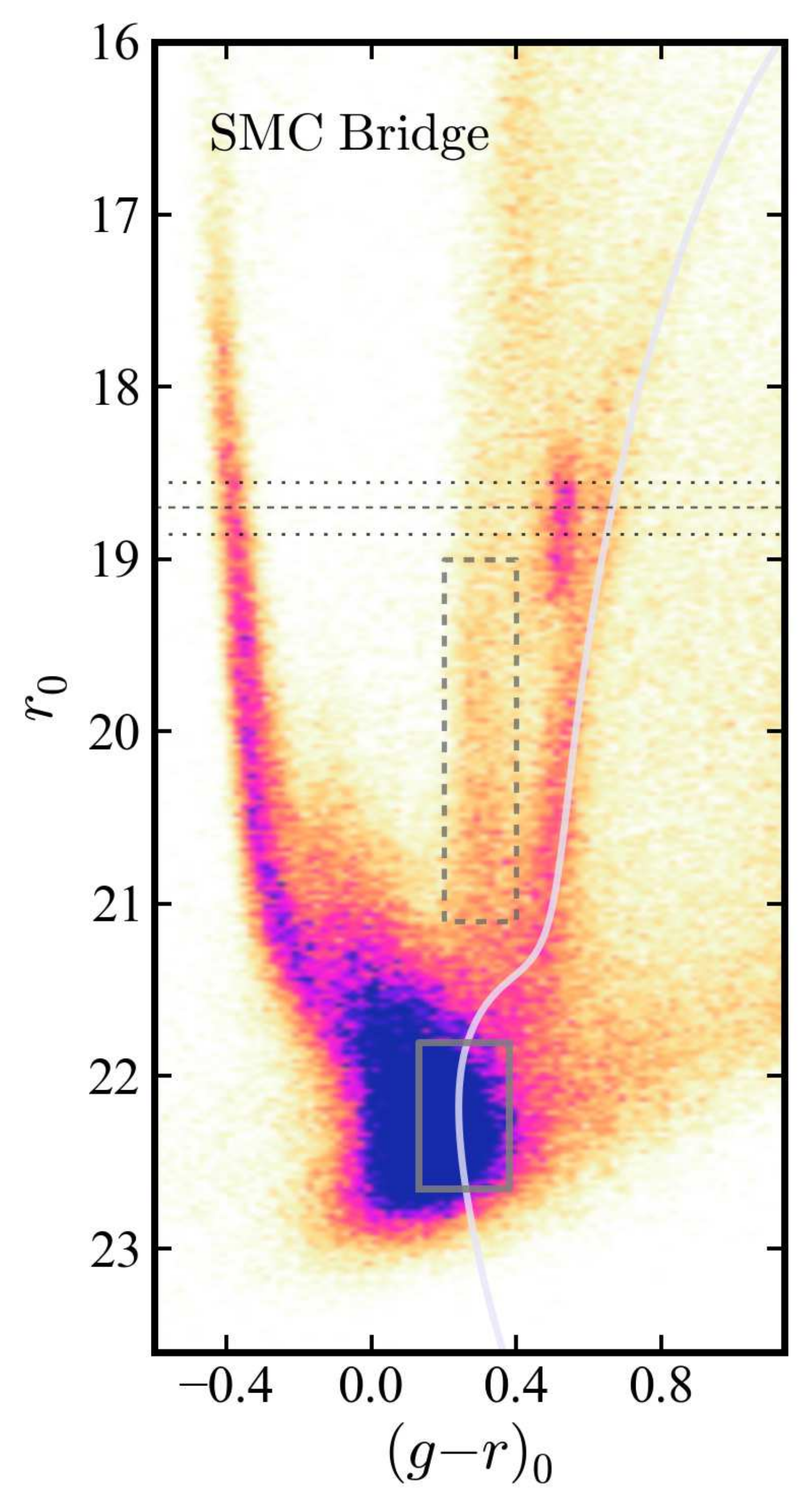}\\
\vspace{1mm}
\includegraphics[height=78mm]{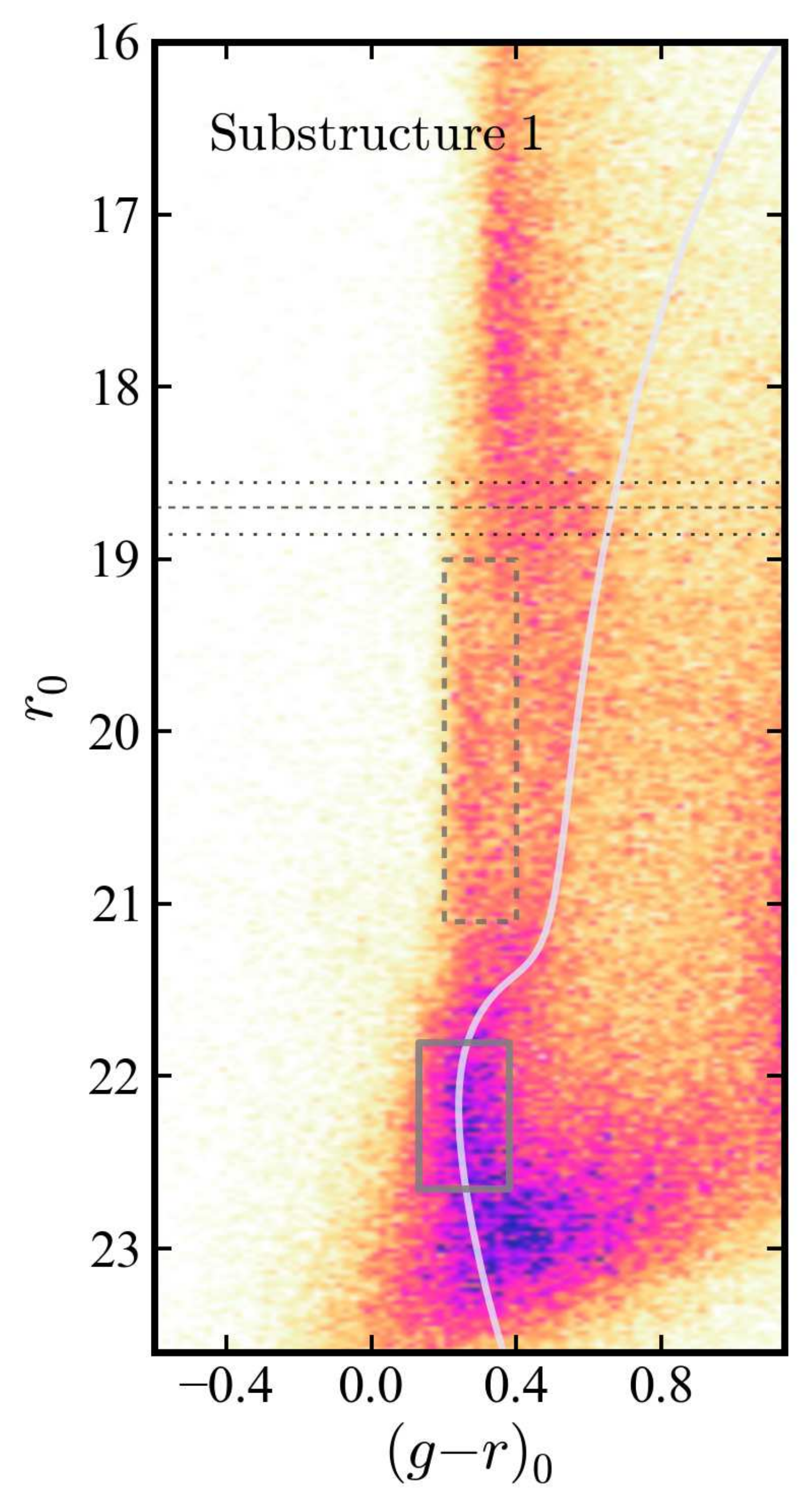}
\hspace{1mm}
\includegraphics[height=78mm]{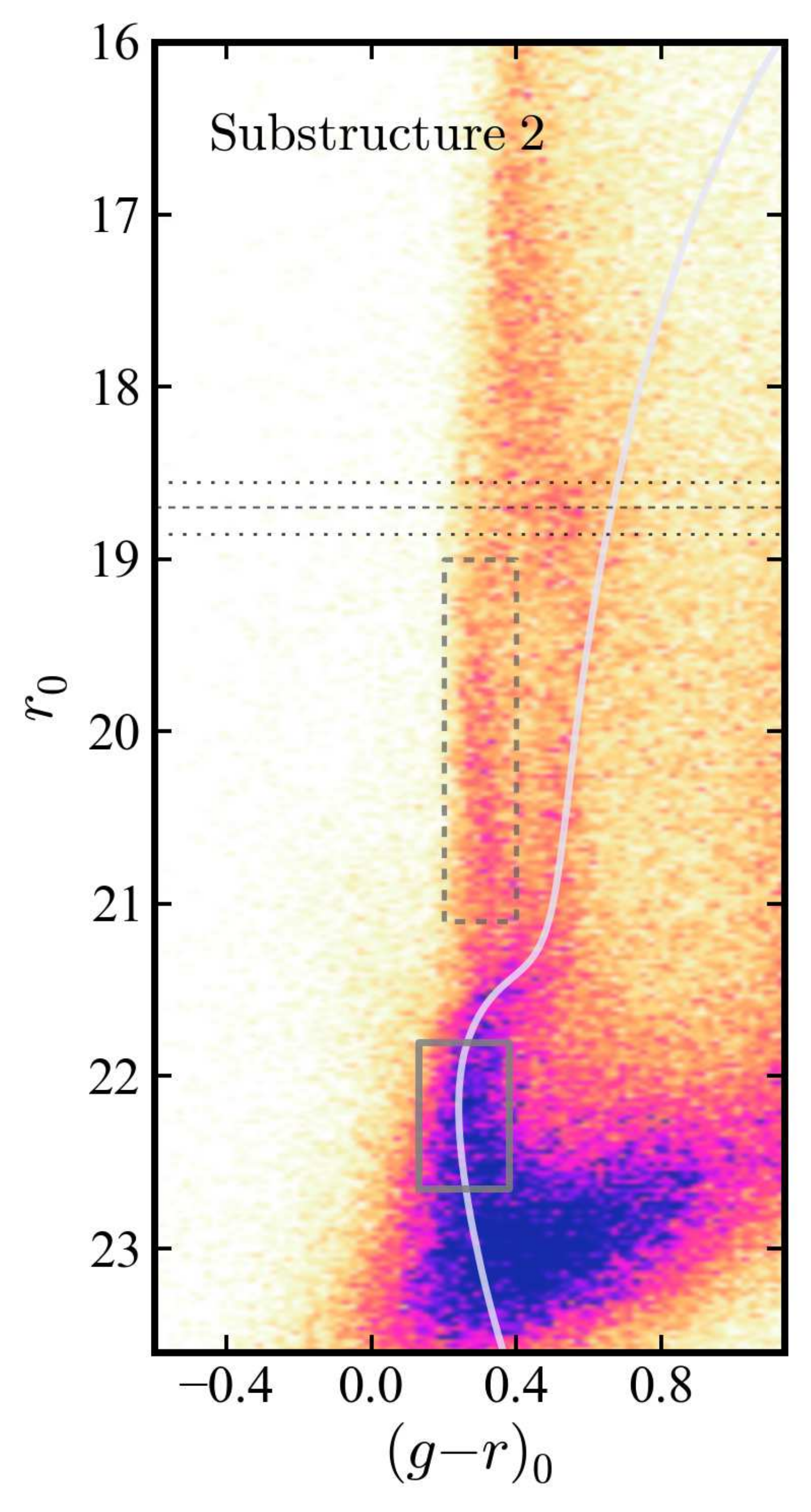}
\hspace{1mm}
\includegraphics[height=78mm]{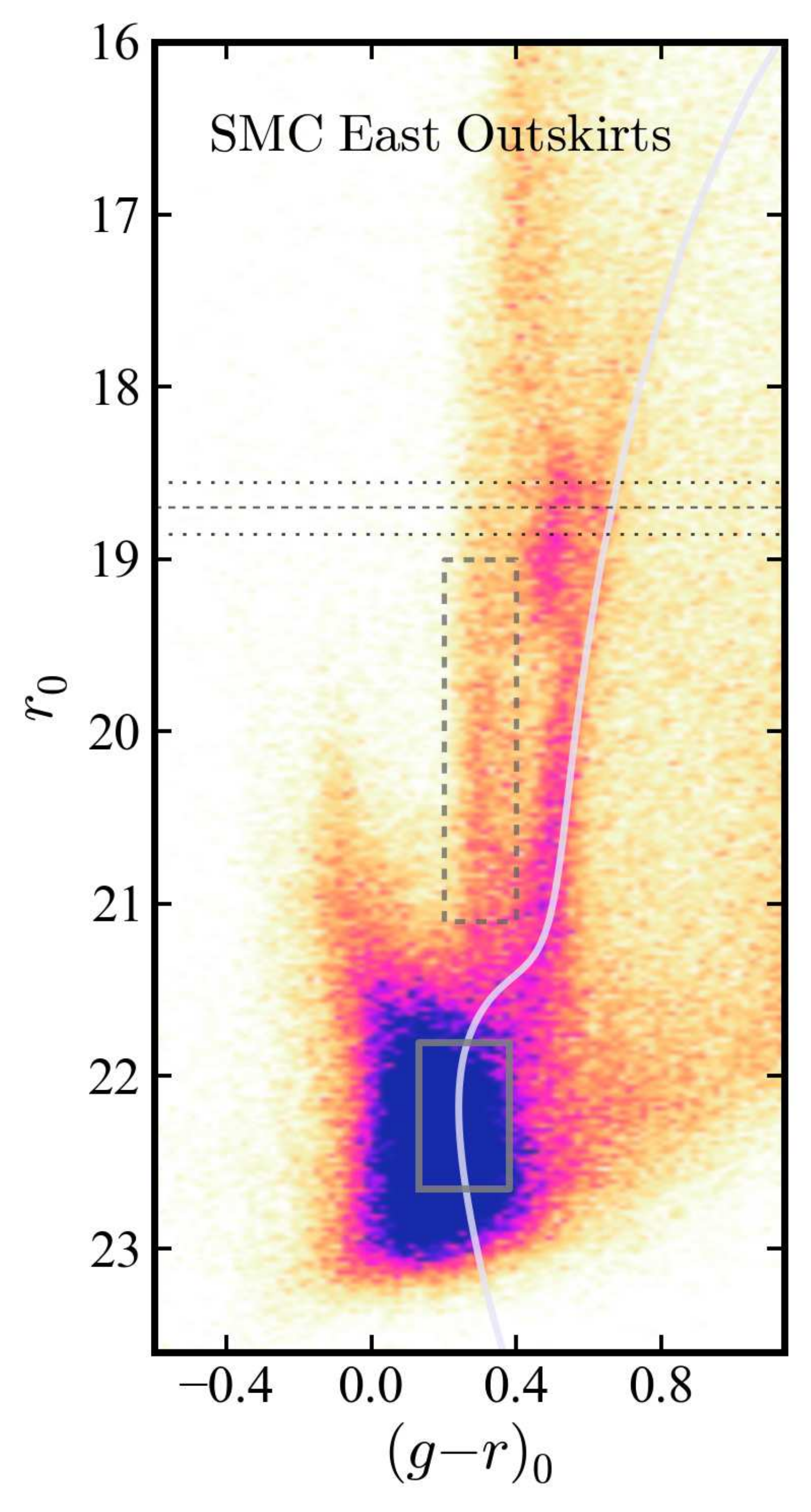}
\hspace{1mm}
\includegraphics[height=78mm]{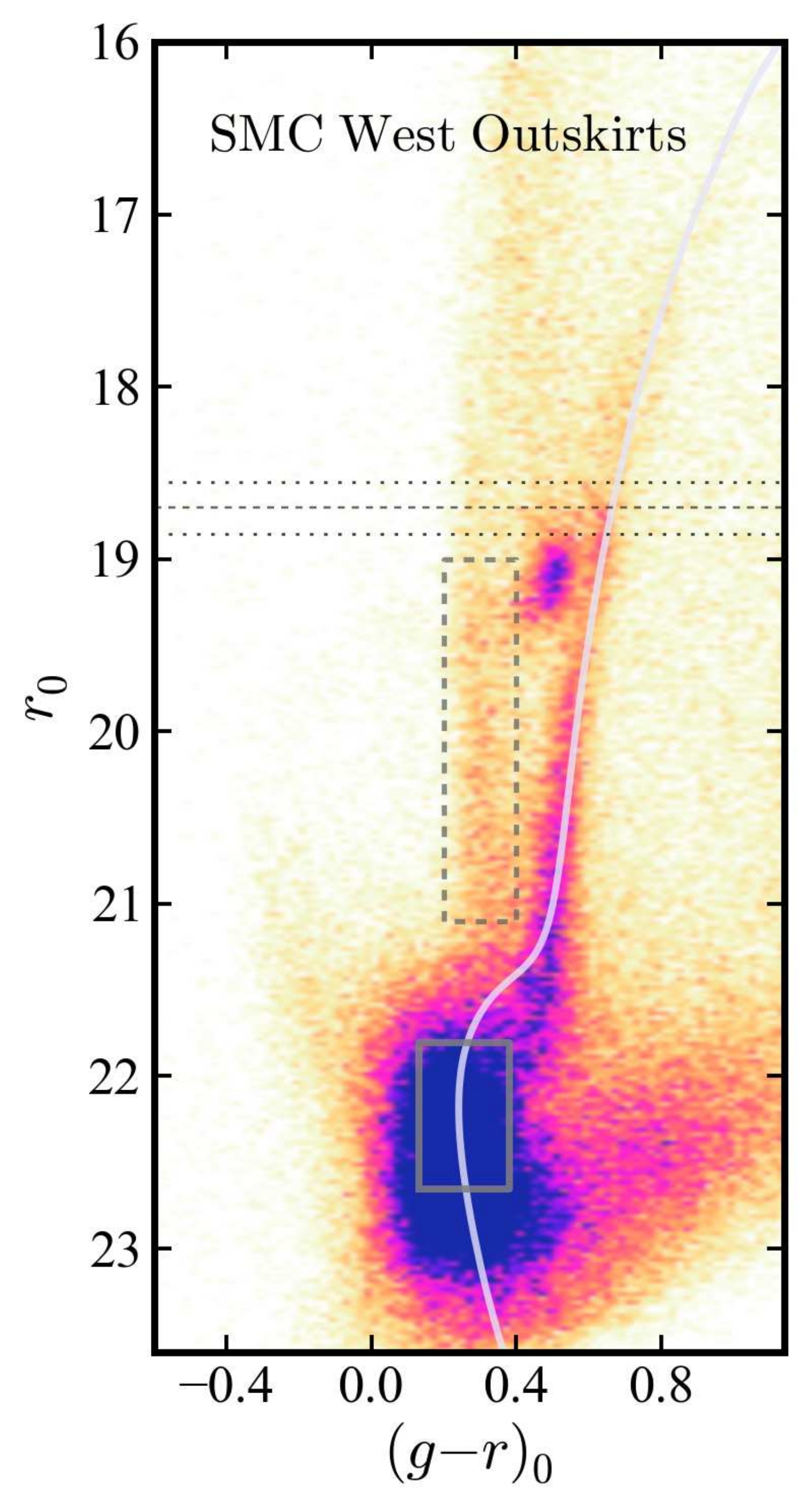}
\end{center}
\caption{CMDs for key regions of interest, as marked on the accompanying map. In each panel the fiducial sequence (white line) and red clump level (dotted black lines) are those derived for the northern LMC disk CMD shown in Figure \ref{f:cmdmap}. 
\label{f:cmds}}
\end{figure*}

 
Figure \ref{f:cmds} shows CMDs for key regions of interest across the Magellanic system. The fiducial sequence and red clump level marked on each CMD are those derived for the northern LMC disk in Figure \ref{f:cmdmap}. We sample the southern LMC disk by selecting stars at position angles $150\degr-210\degr$ east of north, but smaller radii $8\degr-10\degr$ from the LMC centre due to the truncation of the disk in this direction. The southern disk CMD matches the northern fiducial sequence and red clump level well. This confirms the observation by \citet{mackey:16} that the orientation of the line-of-nodes runs almost north-south in the LMC periphery; if the orientation was significantly different, we would expect an offset between north and south of several tenths of a magnitude.\vspace{2mm}

Our CMDs for the two main substructures to the south of the LMC do not have high contrast against contaminating populations due to the low surface density of these features. However, there is no indication that either structure possesses a significantly different line-of-sight distance than the northern and southern LMC disk regions. \citet{belokurov:17} found that the RR Lyrae bridge connecting the Clouds exhibits a substantial line-of-sight depth, including a sizeable component at the LMC distance that stretches at least two-thirds of the way to the SMC. This implies that the structure coincident with the RR Lyrae bridge in projection is likely co-spatial in three dimensions.\vspace{2mm}

The structure of the SMC is known to be highly complex \citep[see e.g.,][and references therein]{nidever:11,nidever:13}, and this is clearly evident in our CMDs. That for the western side of the SMC exhibits a compact red clump sitting $\sim0.4-0.5$ mag fainter than the level observed in the northern and southern LMC disk regions. This offset is also visible at the subgiant branch level, and is consistent with canonical SMC distance estimates \citep{degrijs:15}. In contrast, the red clump for a comparable region on the eastern side of the SMC exhibits a striking vertical extension, which \citet{nidever:13} showed can be almost completely attributed to a large line-of-sight depth. Taking this conclusion at face value, our CMD indicates the presence of some populations in this region that must sit even closer than the northern and southern LMC disk. It is also notable that the eastern side of the SMC exhibits a much more substantial intermediate-age population than at equivalent radii to the west. No comparable intermediate-age population is visible in the CMDs for either of the southern substructures.\vspace{2mm}

The ``SMC Bridge'' CMD in Figure \ref{f:cmds}  encompasses the outer SMC wing that extends into the young stellar bridge stretching between the Clouds \citep[e.g.,][]{irwin:90,battinelli:92,belokurov:17,mackey:17}. The young populations in this region are striking, and are not seen in our other SMC CMDs. However, this part of the SMC also clearly possesses a significant intermediate-age population, and exhibits the same line-of-sight extension seen in the CMD for the eastern SMC outskirts.\vspace{2mm}

\begin{figure*}
\begin{center}
\includegraphics[width=40mm]{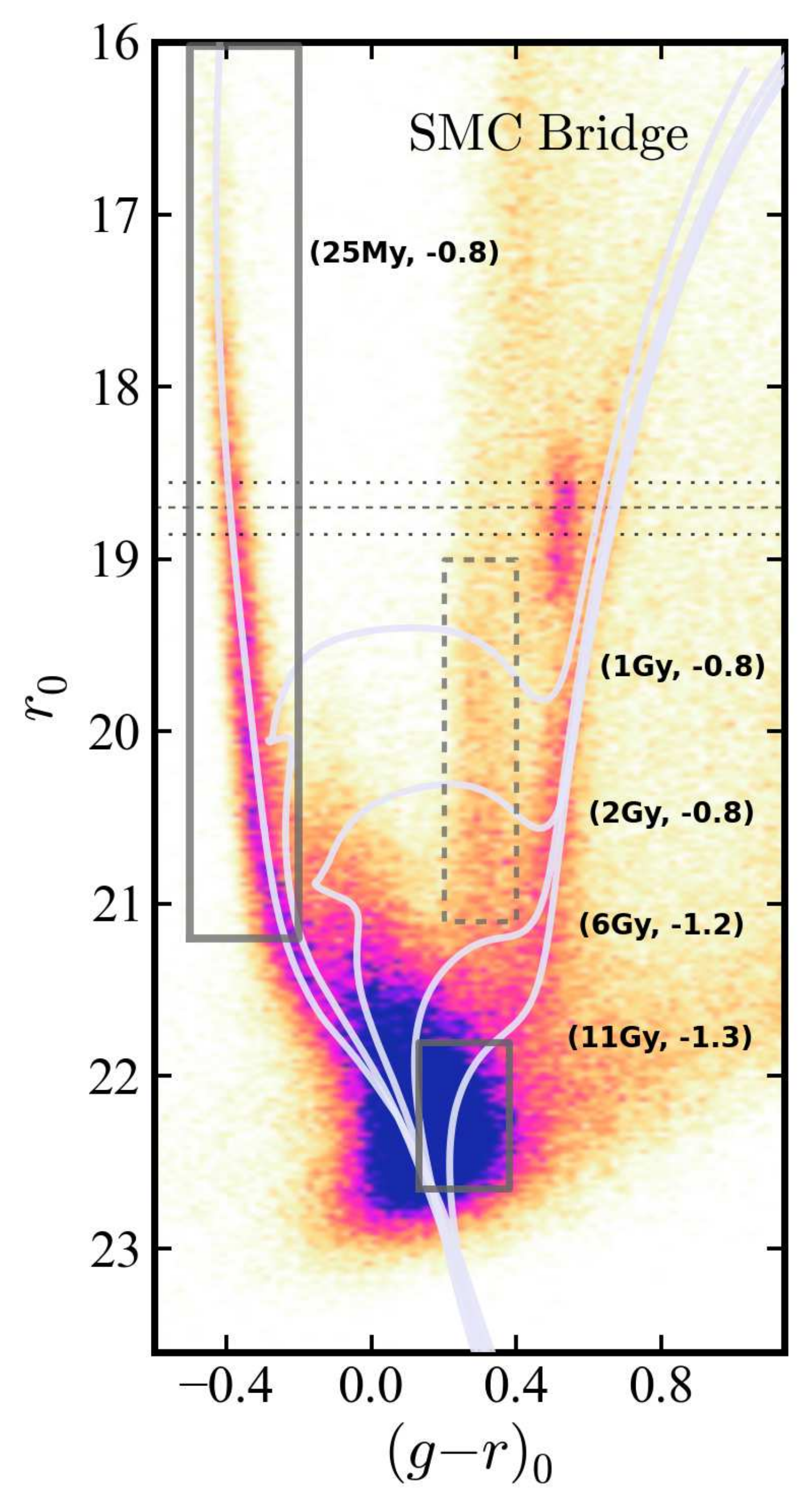}
\hspace{0mm}
\includegraphics[width=130mm,trim=0 -14mm 0 0]{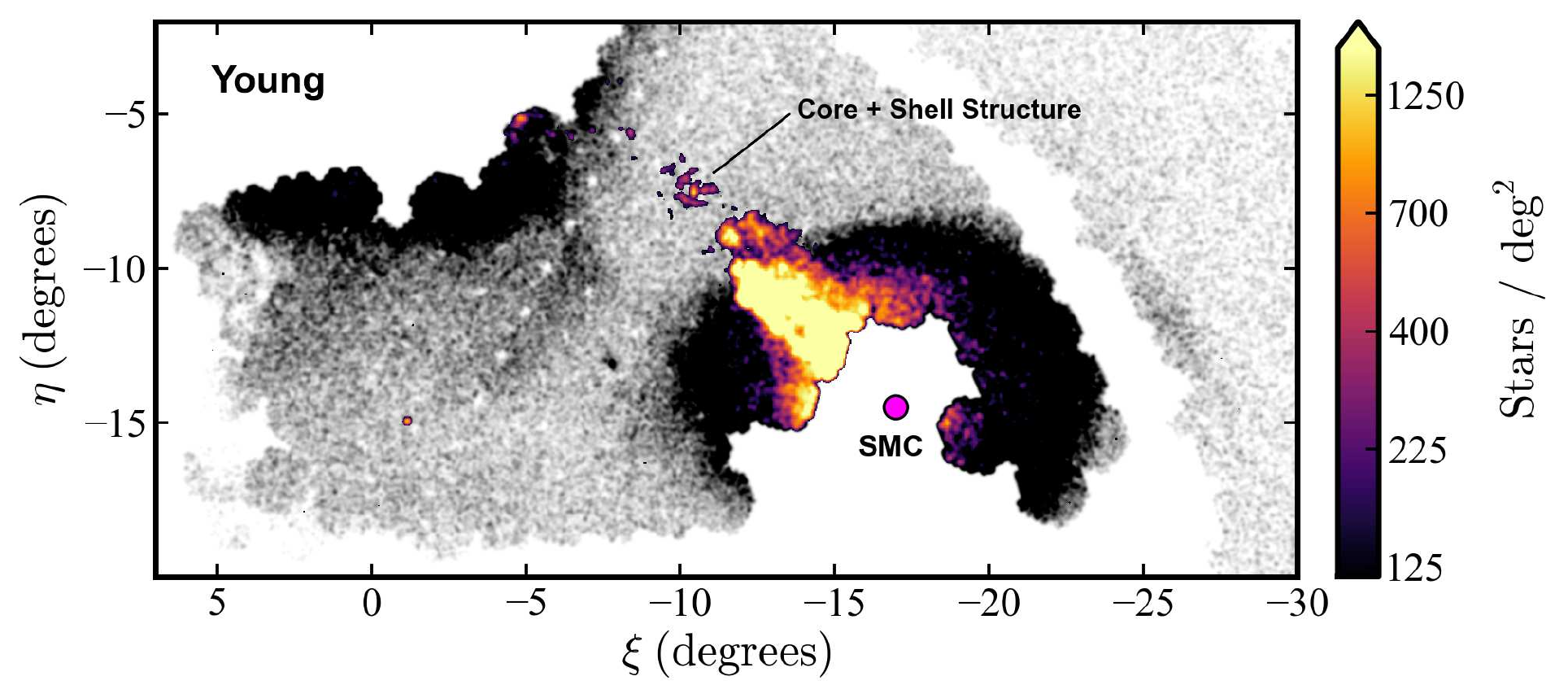}\\
\vspace{1mm}
\includegraphics[width=40mm]{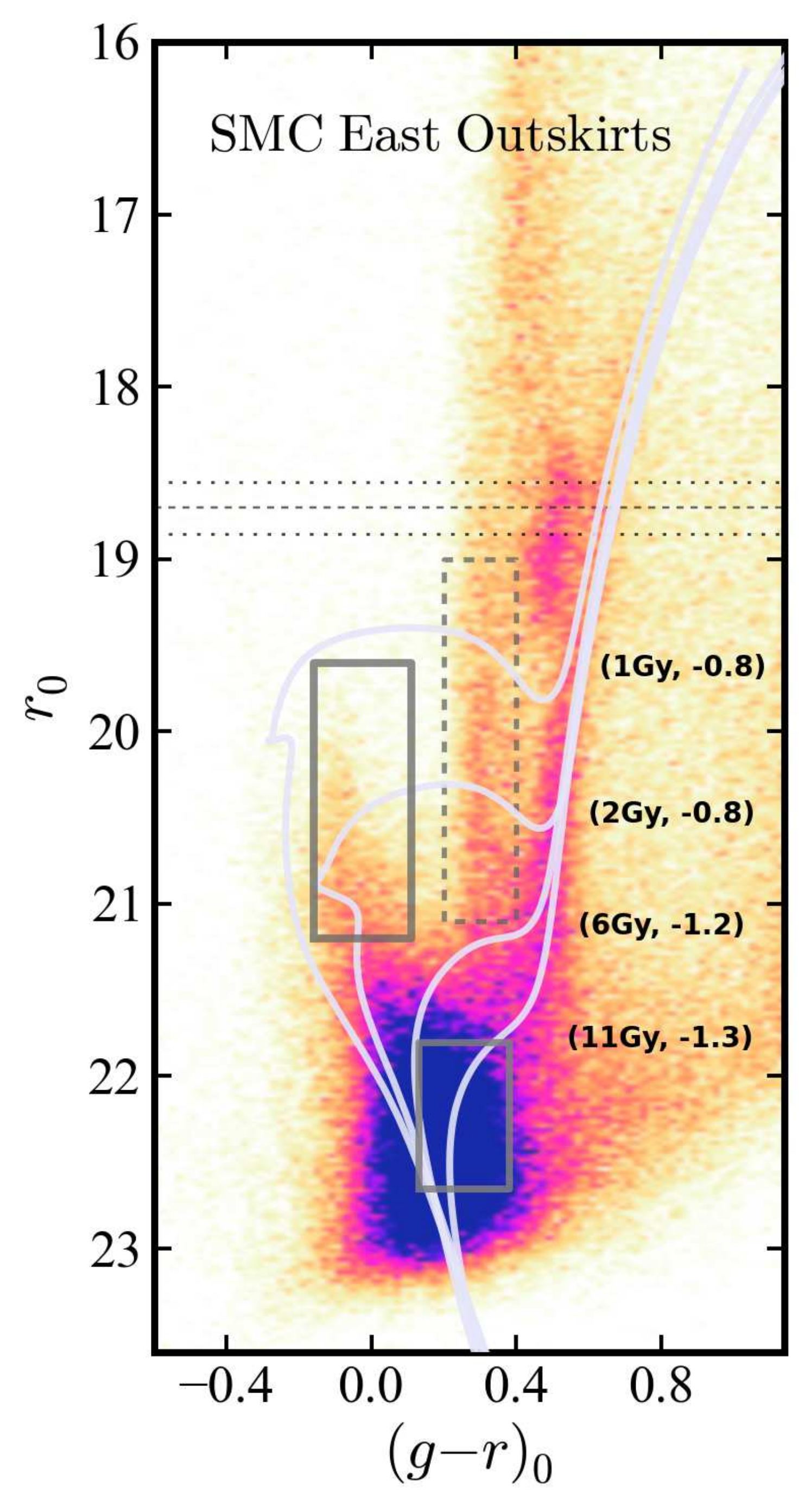}
\hspace{0mm}
\includegraphics[width=130mm,trim=0 -14mm 0 0]{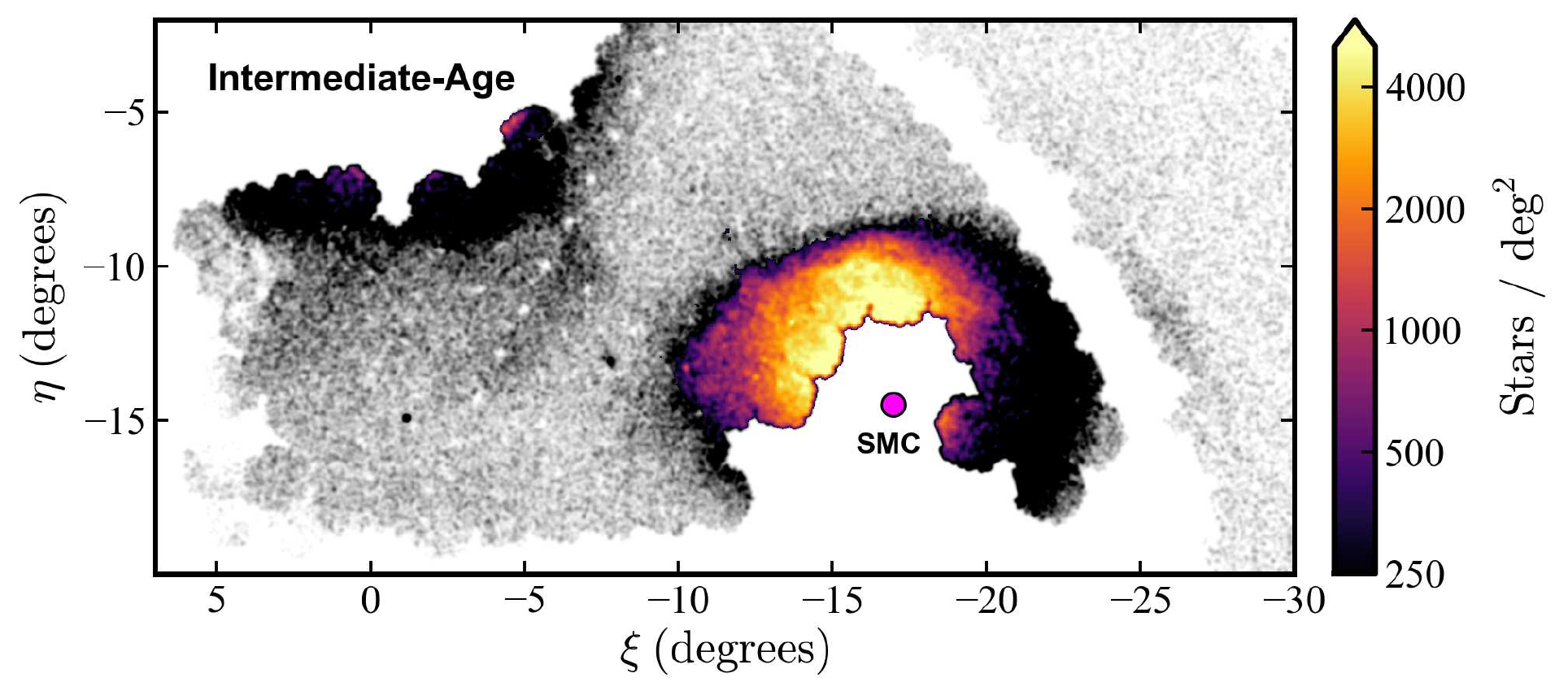}
\end{center}
\caption{Distribution of young and intermediate-age stellar populations across the SMC and inter-Cloud region. We use the SMC Bridge CMD (upper-left) to define a selection box encompassing the sequence for stars with ages $\ll1$ Gyr. The density map for stars in this box (upper-right) clearly reveals the well-known young stellar bridge stretching between the Clouds; the core-shell structure studied by \citet{mackey:17} is labelled. We define an intermediate-age selection box for stars with ages $\sim1.5-4$ Gyr using our CMD for the eastern SMC (lower-left). The map for these stars (lower-right) shows that their distribution is roughly spheroidal with a centroid clearly offset from that for ancient stars. In both CMDs the white lines denote MIST isochrones with age and metallicity as indicated, shifted to $\mu=18.75$. 
\label{f:bridges}}
\end{figure*}

Figure \ref{f:bridges} maps the distribution of young and intermediate-age populations across the SMC and inter-Cloud region. The young sequence evident in the SMC Bridge CMD is well fit by a MESA isochrone of age $25$ Myr and $[$Fe$/$H$]=-0.8$, shifted to a distance modulus $\mu=18.75$. This is approximately the mean line-of-sight distance indicated by the extended red clump. We use this CMD to define a selection box around the young main sequence, and show the spatial density of young stars in the upper-right panel. The outer SMC wing and young stellar bridge are prominent. In the inter-Cloud region the young stars are highly clustered and closely trace the H{\sc i} \citep[e.g.,][]{mackey:17}. This young stellar bridge is offset from the RR Lyrae bridge and old stellar substructures by $\approx5\degr$, likely because of  the ram pressure from the Milky Way's hot corona \citep{belokurov:17}.\vspace{2mm}

To trace the intermediate-age populations we use our eastern SMC CMD to define a selection box that is sensitive to stars of ages $\sim 1.5-4$ Gyr. The spatial density of stars in this age range is quite different from that for either younger or older populations. The intermediate-age stars do not have a significant presence in the inter-Cloud region and instead exhibit a rather spheroidal distribution confined within the SMC. Even though our map does not fully cover the SMC's interior, it is clear that the centroid for this population is offset from that for the old stars by several degrees towards the LMC. A simple comparison of our CMDs for the eastern and western sides of the SMC highlights this lop-sided distribution.\vspace{2mm}

\section{Discussion and Summary}
We have used DECam to map a large portion of the Magellanic periphery and inter-Cloud region to extremely low surface brightness, revealing several key characteristics that provide clues to the recent evolution of the system:
\begin{enumerate}
\item{The outer LMC disk is significantly truncated in both the west and south compared to its extension in the north. The northern and southern regions have similar line-of-sight distances, implying that the apparent truncation is not a perspective effect and, moreover, that the LMC line-of-nodes at radii beyond $\approx9\degr$ must be aligned nearly north-south. This is different from the orientation inferred for inner regions of the LMC by $\sim20\degr-60\degr$ \citep[e.g.,][]{vdm:01,subra:10}, as expected if the LMC disk is warped.}
\item{Large diffuse substructures comprised of ancient stellar populations are present both to the north and south of the LMC, and in the inter-Cloud region. At least one of these features appears co-spatial with the RR Lyrae bridge discovered by \citet{belokurov:17} that connects the Clouds.}
\item{The SMC is strongly distorted, exhibiting (i) twisting of isophotes with increasing galactocentric radius; (ii) tidal tails extending into the RR Lyrae bridge towards the LMC and the \citet{pieres:17} overdensity in the opposite direction; and (iii) a very large line-of-sight depth on the side of the system closest to the LMC.}
\item{Substantial young populations are present in the outer SMC wing, which morphs into a second stellar bridge joining the Clouds. This bridge closely follows the peak of the H{\sc i} distribution in the inter-Cloud region \citep[e.g.,][]{dinescu:12,skowron:14} and is offset from the ancient substructures by $\approx5\degr$.}
\item{Intermediate-age populations are largely confined to the SMC interior, where they exhibit a spheroidal distribution with a centroid offset from that of the old populations by several degrees towards the LMC. This cannot be a perspective effect \citep[cf.][]{nidever:11} as both the intermediate-age and old populations appear to share a similar extension along the line-of-sight.}
\end{enumerate}
While some of these features have been known for years, our deep, panoramic survey has laced them into a coherent picture that starkly illustrates the strongly interacting nature of the LMC-SMC pair. In future, careful characterization of the main structural distortions in the Magellanic outskirts, combined with detailed numerical modeling, holds the promise of providing tight constraints on the allowed orbital parameter space -- in particular, the masses of the two Clouds, the timescale on which they have been bound, and the frequency and impact parameter of recent close passages. For now, we hypothesize that our two new southern substructures could plausibly represent either (i) disrupted regions of the outer LMC disk, or (ii) strong periodic stripping of SMC stars. Since the substructure CMDs do not obviously exhibit the intermediate-age populations so prevalent in the eastern SMC we tentatively favour the former option; future spectroscopic observations should be decisive given the difference in mean metallicity between the Clouds. The strong offset between the intermediate-age and old stellar populations in the SMC is also instructive. Since these stars orbit in the same potential this arrangement must represent formation conditions $\sim1.5-4$ Gyr ago, and we speculate that the gravitational influence of the LMC may already have been affecting the gaseous component of the SMC at this time. If so, our observations imply that the Magellanic Clouds have likely been a bound pair for at least the past several Gyr, as appears probable if the Magellanic system is on its first infall \citep{kallivayalil:13}.\vspace{2mm}

\acknowledgements
ADM holds an Australian Research Council (ARC) Future Fellowship (FT160100206).  ADM and GDC acknowledge support from ARC Discovery Project DP150103294. The research leading to these results has received funding from the European Research Council under the European Union's Seventh Framework Programme (FP/2007-2013)/ERC Grant Agreement no. 308024. We thank the International Telescopes Support Office at the Australian Astronomical Observatory for providing travel support.

This project used data obtained with the Dark Energy Camera (DECam), and public archival data from the Dark Energy Survey (DES). Funding for the DES Projects has been provided by the DOE and NSF (USA), MISE (Spain), STFC (UK), HEFCE (UK), NCSA (UIUC), KICP (U. Chicago), CCAPP (Ohio State), MIFPA (Texas A\&M), CNPQ, FAPERJ, FINEP (Brazil), MINECO (Spain), DFG (Germany), and the collaborating institutions in the Dark Energy Survey, which are Argonne Lab, UC Santa Cruz, University of Cambridge, CIEMAT-Madrid, University of Chicago, University College London, DES-Brazil Consortium, University of Edinburgh, ETH Z{\"u}rich, Fermilab, University of Illinois, ICE (IEEC-CSIC), IFAE Barcelona, Lawrence Berkeley Lab, LMU M{\"u}nchen and the associated Excellence Cluster Universe, University of Michigan, NOAO, University of Nottingham, Ohio State University, University of Pennsylvania, University of Portsmouth, SLAC National Lab, Stanford University, University of Sussex, and Texas A\&M University.

Based on observations at Cerro Tololo Inter-American Observatory, National Optical Astronomy Observatory (programs 2016A-0618 and 2017B-0906, PI: Mackey), which is operated by the Association of Universities for Research in Astronomy (AURA) under a cooperative agreement with the National Science Foundation.\vspace{2mm}

\facility{Blanco (DECam)}

\end{document}